\documentclass[journal]{IEEEtran}

\usepackage{cite}
\usepackage{algorithmic}
\usepackage{makecell}

\usepackage{color,soul}
\setulcolor{red}
\sethlcolor{yellow}
\renewcommand\hl[1]{#1} 

\usepackage{etoolbox}
\apptocmd{\sloppy}{\hbadness 10000\relax}{}{}
\usepackage{amsmath,amssymb,amsfonts}
\ifCLASSINFOpdf
  \usepackage[pdftex]{graphicx}
\else

\fi

\usepackage[dvipsnames]{xcolor}
\usepackage[caption=false]{subfig}
\usepackage{url}
\usepackage{bm}

\usepackage{tabularx, booktabs}

\begin{document}
\bstctlcite{IEEEexample:BSTcontrol}

\title{Closing the Management Gap for Satellite-Integrated Community Networks: \\A Hierarchical Approach to Self-Maintenance}

\author{Peng~Hu,~\IEEEmembership{Senior Member,~IEEE}
}

%

\markboth{\copyright IEEE 2021. Published In: IEEE Communications Magazine, vol. 59, no. 12, pp. 43-49, December 2021, doi: 10.1109/MCOM.001.2100479}
{}



\maketitle

\begin{abstract}
Community networks (CNs) have become an important paradigm for providing essential Internet connectivity in unserved and underserved areas across the world. However, an indispensable part for CNs is network management, where responsive and autonomous maintenance is much needed. With the technological advancement in telecommunications networks, a classical satellite-dependent CN is envisioned to be transformed into a satellite-integrated CN (SICN), which will embrace significant autonomy, intelligence, and scalability in network management. This article discusses the machine-learning (ML) based hierarchical approach to enabling autonomous self-maintenance for SICNs. The approach is split into the anomaly identification and anomaly mitigation phases, where the related ML methods, data collection means, deployment options, and mitigation schemes are presented. With the case study, we discuss a typical scenario using satellite and fixed connections as backhaul options and show the effectiveness and performance improvements of the proposed approach with recurrent neural network and ensemble methods.
\end{abstract}


\IEEEpeerreviewmaketitle

\section{Introduction}

\IEEEPARstart{T}{he} satellite networks have long become a key connectivity option for community networks (CNs) \cite{Micholia2018} in unserved and underserved areas worldwide. With the international efforts on closing the ``broadband gap'' for digital divide including the implementation of new satellite network and telecommunications infrastructures in rural and remote areas, the traditional satellite-dependent CNs (SDCNs) are envisioned to be transforming into a satellite-integrated CN (SICN), featuring an integration of heterogeneous networks and segments to provide broadband, resilient, and agile end-to-end connections. However, this transformation imposes unprecedented challenges to CN management centered on the assurance of critical operation together with low network management costs, high responsiveness and scalability. These challenges are often perceived as barriers and implicitly create a ``management gap'' for CNs. To close this gap, the SICN calls for self-maintainability leading to a responsive, autonomous, and scalable management solution. Such self-maintenance capability will also greatly enhance the critical role of CNs in supporting various applications for education, businesses, facilities, environmental monitoring, the Internet of Things, etc. 

Self-maintenance in network management is an important task for SICNs, which requires autonomous identification, planning, and execution for fixes and upgrades of network resources. Today's machine learning (ML) methods provide a solid foundation for realizing self-maintenance for CNs. The overall use of ML on network management has been underpinned by the recent efforts from standards development organizations. The latest 3GPP Rel. 17 has provided two management architectures for integrated satellite components with a 5G network. The ITU Telecommunication Standardization Sector (ITU-T) has identified the gaps to achieve the network 2030 goals and the architectural framework for ML in future networks. The European Telecommunications Standards Institute (ETSI) has created an industry specification group for the zero-touch network and service management, where largely autonomous networks without further human intervention are envisioned goals with the proposed means of network automation. However, specific solutions to self-maintenance for CNs have not been available in these efforts.

Current ML methods are centered on network intrusion detection (NID) for cyberattacks, small-scale or special network setups, and anomaly detection using telemetry data on satellite and network devices \cite{Pacheco2020}. The heterogeneity and complexity of connectivity options, architectures, and anomalous network events, including the high-impact low-frequency (HILF) incidents, will further make the ML-based solutions challenging on SICNs. For example, SICNs tend to use various architectures, technologies and scales. The expected use of new Geosynchronous Equatorial Orbit (GEO) and non-geostationary (NGSO) satellites requires new satellite-terrestrial integrated network setups. With various technologies used in the access networks for geographically distributed communities, using ML methods in an integrated fashion with a combination of fixed, wireless, and satellite networks requires an approach with holistic thinking. This approach can accommodate additional entities on SICNs such as data centers (DCs) and Internet exchange points (IXPs), where these entities may be inconsistent across remote communities \cite{Waites16} and introduce anomalous events. The support for current management data and platforms that are currently segmented and underutilized should be considered within the approach.

This article serves two purposes. First, we present the SICN and its management gap in self-maintenance. Second, we propose an ML-based hierarchical approach for self-maintenance solutions with anomaly identification and mitigation phases. The ML models in a typical scenario are analyzed, and the ensemble and deep learning models are proposed for the anomaly identification phase. The proposed approach aims to:
\begin{itemize}
    \item Address the evolving connection diversity on SICNs while maintaining compatibility \hl{with SDCNs}.
    \item Bridge the gap between network anomaly identification and mitigation measures, and integrate the NID, fault detection and system reliability.
    \item Leverage various datasets and reduce learning time in computation without performance compromise.
    \item Enable multi-scale root cause analysis of anomalous network events.
\end{itemize}

\begin{figure}[!ht]
\centering
\frame{\includegraphics[width=0.74\linewidth]{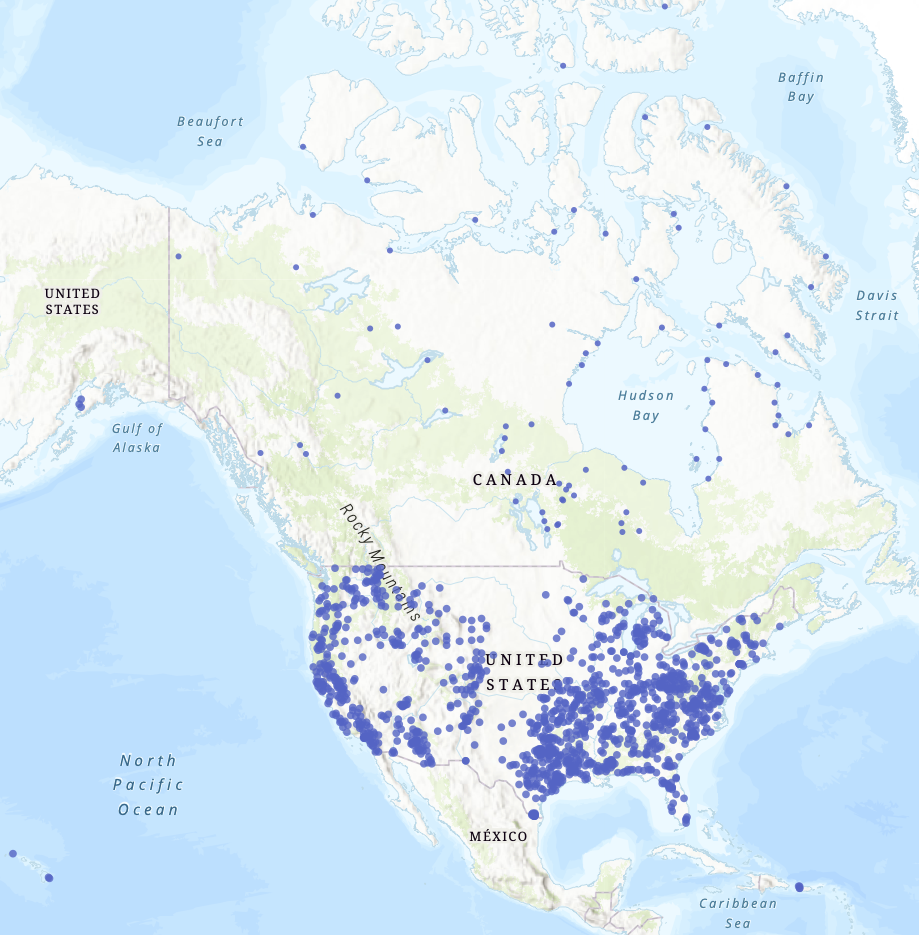}}
\caption{Satellite-dependent communities (labeled as blue dots) in the US and Canada. (Only communities with greater than 99 households are plotted to keep the consistency of the magnitude of datasets.) Data source: 2020 National Broadband Data Canada and ACS Internet Connectivity Data}
\label{Fig:map}
\end{figure}

\begin{figure*}[!ht]
\centering
\includegraphics[width=0.82\linewidth]{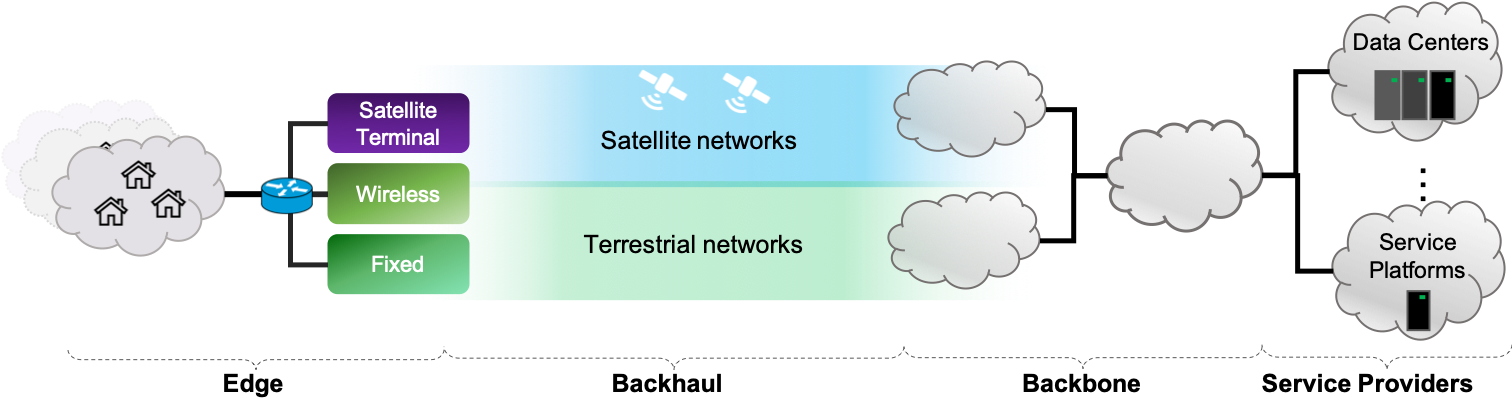}
\caption{A generic architecture of community networks.}
\label{Fig:arch_cn}
\end{figure*}

\section{Moving from Satellite-Dependent to Satellite-Integrated CNs}

The concept of CNs emerged in the late 90s as indicated in \cite{Micholia2018} and then evolved into two general options for bringing Internet access to unserved and underserved areas: using available and economic tools/infrastructures, or providing broadband access under the umbrella of ``alternative networks'' \cite{saldana2016alternative}. Satellite networks have been a key connectivity option to the CNs. In the Broadband Coverage in Europe report 2017, satellite broadband is considered ``the most pervasive technology in Europe in terms of overall coverage''. In the US and Canada as shown in Fig. \mbox{\ref{Fig:map}}, there are also a large number of SDCN communities and users geographically distributed across the countries.

SDCNs are expected to be transformed into satellite-integrated CNs (SICNs), aligned with recent developments in NGSO satellites such as LEO satellite mega-constellations and international/regional development plans for the terrestrial Internet infrastructure. The integration of satellite and telecommunications networks has been discussed in the literature under the umbrella of space information networks, space-terrestrial integrated networks, or space-air-ground integrated networks, where the LEO satellites and high-altitude platform (HAP) components can be added for realizing backhaul links. Although these new paradigms provide relevant discussions to SICNs, they have hardly addressed the self-maintenance topic. While keeping the compatibility with SDCN, the new SICN paradigm will make the traditional SDCN users be able to access to the Internet through advanced satellite networks with fixed and/or wireless infrastructures, where the satellite can either be a backhaul link or supplementary to the backhaul options provided by terrestrial networks. In this context, the SICN will also include CNs which are currently non-satellite-dependent and will add satellite backhauls to their existing connectivity options.

An example architecture of an SICN is shown in Fig. \mbox{\ref{Fig:arch_cn}}, where there are edge, backhaul, and backbone segments between CN users and service providers. The CNs are connected to the network edge with at least a satellite frontend, backhaul links provided by satellite and/or terrestrial networks, and the backbone of the Internet. The edge and backhaul portions represent the most possible places \hl{requiring} maintenance efforts\hl{.} The broadband access technologies \hl{shown in Fig. \mbox{\ref{Fig:arch_cn}}} can be categorized into fixed networks, including copper/cable and fiber optic options, wireless networks including cellular and microwave options, and satellite networks\hl{.}

\section{Self-Maintenance for SICNs}
Self-maintenance is a new design methodology in telecommunications systems. An earlier discussion regarding self-maintenance was rooted in manufacturing systems, which is viewed as a capability to monitor and diagnose itself and maintain its functions in case of failures or degradation. The related discussion has been seen in the recent ETSI White Paper No. 40 for self-healing and self-monitoring capabilities of autonomous networks (ANs). Some elements of self-maintenance stem from the prediction methods, which have been used in telecommunications networks. It is generally related to self-management mentioned in ``zero-touch management'' and ``autonomic networks'' within ETSI and the Internet Engineering Task Force (IETF), respectively.

Although satellite networks have become an important way for closing the broadband gap, there exists an equally important management gap. In fact, in addition to broadband access, CN users often suffer from poor Internet speeds, high connectivity costs, and various network interruptions. For the SICN users such as the communities shown in Fig. \ref{Fig:map}, the network performance is dependent on various factors on the access/edge network and satellite backhauls.

\begin{figure*}[!ht]
\centering
\includegraphics[width=\linewidth]{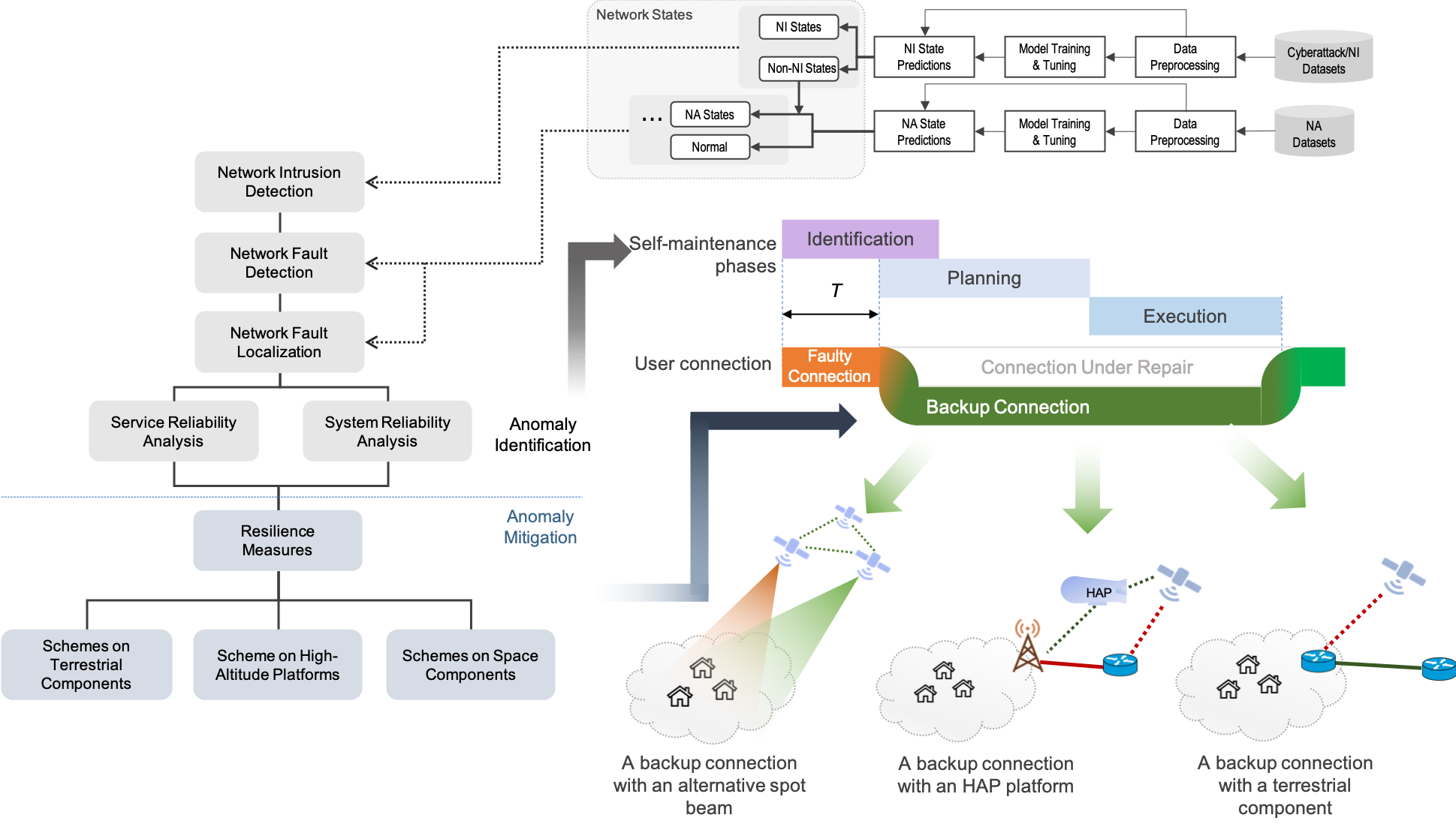}
\caption{\hl{A hierarchical approach to self-maintenance for SICNs.}}
\label{Fig:process}
\end{figure*}

The important step in achieving self-maintainability is identifying the causes to anomalous network events. As an example \hl{process shown} in Fig. \ref{Fig:process}, once the faulty connection occurred on a user connection, the identification phase starts and a mitigation scheme is executed as part of the planning phase, where a backup connection can be scheduled before the connection repair is done through the execution phase. The time $T$ taken in the identification phase is essential for improving user experience, where the less the value of $T$ the better\hl{. For} an SICN, anomalous events may come from different network segments, the connections between them, and architectural entities such as an IXP or DC in between. However, there are a number of causes contributing to the anomalies. For example, within a satellite network, the anomalies \cite{Galvan2014} may come from the space environment (e.g., cosmic rays, the Van Allen radiation belts, etc.), mission degradation on subsystems/services, electrostatic charge, operation errors, and malicious actions. For LEO satellite constellations, these errors may result in faulty inter-satellite link (ISL) issues. The \hl{atmospheric} conditions may contribute to the \hl{influence of the} satellite-ground radio-frequency (RF) links.
The failures on other network segments may result from the device hardware failures on network nodes, such as processors, storage devices, power modules, and network interfaces. These failures occurred on a single compute node can cause various outage events, and 
ultimately affect network status or metrics, such as link state, packet loss, latency, throughput, and congestion. On the other hand, software failures from application endpoints will often result in abnormal application traffic. \hl{A} self-maintenance \hl{process} in the execution phase may initiate network automation tools to re-deploy the application services as \hl{a mitigation} scheme.

\section{A Hierarchical Approach}

To enable self-maintenance capability in SICNs, we propose an ML-based hierarchical approach. The proposed approach is depicted in a diagram shown in Fig. \ref{Fig:process}, where the blocks represent the steps in two \hl{general} phases, i.e., anomaly identification phase and anomaly mitigation phase. In the first phase, the NID is done using the cyberattack datasets, and then the detection and localization of non-cyberattack anomalous events will be done to infer more specific causes. \hl{The} root causes can be further identifiable through the service or system reliability analysis, \hl{which} can be traced back to a level of the software service, device, or hardware component. \hl{These} causes located will facilitate the planning and execution of self-maintenance activities, such as scheduling hardware/software fixes and dispatching repair resources from a network operations center (NOC)\hl{. }In the second phase, the resilience measures are employed for mitigating outage links or malfunctioning devices, while \hl{on} an SICN such resilience measures can be implemented \hl{through} space, air or ground components such as redundant satellite channels, HAPs, or base stations, \hl{with possible} redundancy or fail-safe considerations. An example resilience measure using a HAP is discussed in \cite{Hu21}, where HAP entities considering unmanned aerial vehicles (UAVs) or balloons can be dispatched to mitigate link outage events between satellite and terrestrial components. The temporary links provided by HAPs can continuously enable Internet access for community users in an SICN while meeting the key performance metrics.

The hierarchy \hl{of the proposed approach} implies possible types of anomalies in cyberattacks and faulty events, where there are several differences between them. \hl{First,} detecting cyberattacks is often used in \hl{NID} systems that do not cover the faulty events of a network caused by, for example, device malfunctions, interface issues, or link outages. Second, \hl{NID} is usually linked to attack countermeasures, not resilience measures. Third, \hl{NID} can be deployed separately from other steps in Fig. \ref{Fig:process}. In this sense, existing NID systems or deployments can be leveraged. Last but not least, the separation between NID and fault detection is important as NID systems would need to handle a broad attack surface and may impose different requirements for data and compute resources than those in the subsequent steps.

\subsection{Employing ML Methods}

Rule-based reasoning under the umbrella of expert systems \cite{Fuller1999} has been a \hl{traditional} approach to fault management where human experience is modelled with a set of condition-action rules. A codebook approach was considered superior to rule-based reasoning, where a knowledge base for identifying network failures with system events labeled is created as correlation matrices. These approaches can model the human experience or domain knowledge but they lack flexibility as it is often difficult to create a codebook or a set of production rules for a network. Recent ML-based approaches mitigate the requirements for explicitly modelling the managed network with domain knowledge. A random forests \hl{(RF)} algorithm was proposed \cite{Gonzalez2017} to detect the network failures for an industrial network setting based on the features from the device interface and virtual machine (VM) status, while the features and network sizes are not applicable to SICNs. A combined use of Bayesian networks and case-based analysis has been used in identifying the virtual private network (VPN) issues \cite{Bennacer2015} but the network size is small. The border gateway protocol (BGP) data on autonomous systems (ASes) are introduced in \cite{li2019machine, Lopes2019} for anomaly detection. ML-enabled automation in network service management has been presented in \cite{Turk19}, where service quality states on a small network with six routers are projected using decision tree and gradient boosting algorithms. However, the current ML models target limited network resources and datasets on a special and small-scale network, which can hardly be applied to SICNs requiring high accuracy performance with efficient executions. \hl{The hierarchical thinking in ML has emerged only recently. A bi-level hierarchical classification methodology using ML has been proposed in \mbox{\cite{Osman2019}} to identify the different types of secondary tasks drivers are engaged in using their driving behavior parameters. The authors in \mbox{\cite{Alin2020}} indicated that splitting the classification task into sub-classification tasks can improve the accuracy rate on some non-BGP benchmark cyberattack datasets. Currently, ML-based self-maintenance solutions for SICNs are not available.
}

In the proposed approach, ML-based multi-class classification can be formulated as a unified way of handling \hl{NID}, network fault detection/localization, and service/system reliability analysis for the anomaly identification phase. \hl{The} features extracted from the network \hl{flows can be used} to predict the \hl{network states, where multiple network anomalies caused by different factors can be considered, and} the class labels \hl{represent the states}. \hl{The states can be} normal and abnormal states \hl{and extensible to} multiple fine-grained states\hl{. The determination process of root causes to network anomalies following the hierarchical steps can be depicted in the ML pipeline shown on the top right of Fig. 3, where the cyberattack or network intrusion (NI) datasets and network anomaly (NA) datasets from various sources can be utilized in the steps in the anomaly identification phase.}

\subsection{Datasets with Management Tools and Platforms}

The proposed approach can utilize various datasets that can be obtained through different platforms. Overall the data collection efforts can be performed with the network management activities, which are usually done at NOCs with a team of staff members in a telecommunications organization. In a modern NOC that may manage enterprise networks, data centers, and service providers altogether, the management data can also be acquired from multiple network segments. These data can include the existing diagnostic data using Internet control message protocol (ICMP)\hl{, and the} data made available with the simple network management protocol (SNMP) \hl{setups to} facilitate fault detection and diagnosis for an SICN\hl{. The} SNMP's management information base (MIB) can include various management objects (MO) for monitoring managed resources\hl{. The} latest SNMPv3 enhances the security features and can be used to set up a data collection model for proactive monitoring of hardware resources. 
\hl{With a software-defined networking platform}, management data can be acquired with centralized controllers and functions.
Moreover, analyzing the traces of network protocols provides another way to obtain data-driven insights \hl{into network} issues. For example, BGP as an important protocol has been used to reveal the misconfigurations of the networks for network operations. \hl{As} BGP maintains the routes of ASes of network service providers while provides a number of path parameters, \hl{it} can be used to detect faults or anomalies. BGP has also been widely used in modern data centers following the Clos topology.

The existing platforms or protocols provide means for collecting datasets for ML-based solutions, but they \hl{do not} directly support out-of-the-box solutions to the self-maintenance needs for SICNs. The ML-based hierarchical approach is therefore essential to guide the data collection efforts and efficiently provide analytical results for self-maintainability.

\subsection{Discussion on Deployment Options}

The proposed approach can be deployed on top of underlying network architectures and infrastructures. For example, as shown in Fig. \ref{Fig:arch_cn}, we can use an intelligent gateway (IGW) \hl{on} an edge router close to the satellite terminal to implement self-maintenance capability for an SICN. \hl{This IGW} can \hl{perform tasks}, such as (a) making adaptive decisions on switching satellite links between Ka and C bands at the physical layer according to weather conditions, (b) detecting anomalies based on the network traffic and management data, (c) identifying malfunctioning network devices or components based on the management data from the adjacent entities\hl{, (d)} responsively identifying the causes of network interruptions in the access network or beyond with assistance from the adjacent IGW entities, and (e) dispatch HAPs as a resilience scheme to fix link outages. More specifically, when identifying the network issues, the BGP and/or IPFIX/NetFlow data can be utilized at the IGW entities. In case of a network interruption, these entities can help progressively locate the root causes. BGP data \cite{li2019machine} \hl{can be used} to identify the cyberattack-related anomalies in contrast with normal \hl{states}. The IGW entities with the support \hl{from} a local IXP or DC \cite{Lopes2019} can have various BGP event data collected and analyzed with proper ML models\hl{ to }quickly identify the network issues and locate the faulty parts with high accuracy.

There are development variations following the approach. For example, when a network service provider has separate NID entities close to the backbone segment as shown in Fig. \ref{Fig:arch_cn}, the NID tasks \hl{on} the IGW end close to the edge segment can interact with the remote NID \hl{entities} and thus offload the computation at the NID step. The interactions between NID entities can still be considered at the NID step in the anomaly identification phase. \hl{The use of an IGW intends to cause minimum physical changes to the existing networks for an SICN and provide consistent service to users. The IGW can be implemented as software entities running on existing network devices interfacing with satellite and terrestrial networks and with additional application services. IGW can also interact with various data collection entities, such as SNMP-based telemetry data collectors on the existing networks using separate management platforms to facilitate network fault localization. The IGW can coordinate to execute a resilience measure and provide a bridged connection for users during a faulty connection. With the monitoring of link quality and atmospheric parameters for space-ground connections, the IGW module could access a satellite network on a reliable channel optimally.} 
In this article, we focus on the case of an economic and provider-neutral deployment option, where anomaly identification tasks are passively completed on the edge segment.

\begin{figure*}[!ht]
\centering
\includegraphics[width=4.8in]{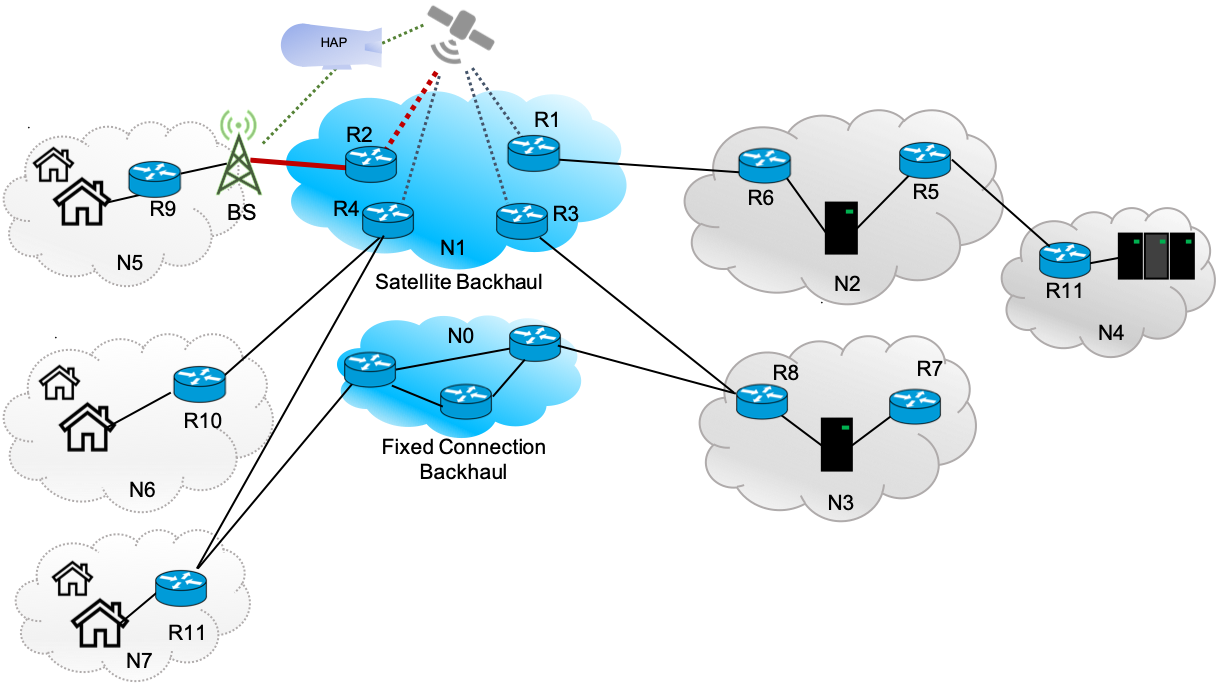}
\caption{An example setup of an SICN.}
\label{Fig:case_study}
\end{figure*}

\section{Case Study}
We apply the ML-based hierarchical approach to a self-maintenance scenario on an SICN shown in Fig. \ref{Fig:case_study}, where satellite and fixed connection backhauls (labeled as N0 and N1) are connected to an SICN including three communities represented by local networks \hl{N5-N7, }and the backbone/service provider networks \hl{N2-N4} are shown on the right. N5 represents an \hl{SICN} setup with cellular distribution, N6 represents the classical SDCN setup, and N7 represents another SICN setup with connection diversity provided by satellite and fixed connections. N7 is connected to both N0 and N1 while N5 is connected to the cellular base station (BS), which \hl{is} connected to the satellite terminals on R2. This setup is representative and can be \hl{scaled up }
where additional network segments can be added on both ends of the satellite backhaul.

To demonstrate the effectiveness\hl{ of }the ML methods, we employ the benchmark BGP datasets\hl{, }where BGP plays an important role in maintaining connectivity on network segments and gateways \hl{on} an SICN. The SICN in Fig. \ref{Fig:case_study} is set up in an emulated network, where \hl{satellite entities} and routers are based on the Mininet virtual machines. Edge routers have been added to each network as an AS to generate and log BGP traffic. The traffic flows from the service providers to CN end-users through N1 and N0. The FRRouting \hl{protocol stack is used to configure BGP, and the external and internal BGP protocols are running between ASes and within an AS.} BGP route information is periodically shared between routers as IGW entities and is stored in dump files. BGP datasets are logged when \hl{routers} advertised their prefixes every few minutes. The data in dump files are then \hl{preprocessed} by the Zebra dump parser and converted into tabular form \hl{for feature extraction} used in \cite{li2019machine}. 

With this generic SICN setup, we consider link outage as a representative type of anomalous network events. Here a link outage may have resulted from cyberattacks, adverse weather conditions on the satellite backhaul link, and a number of device-specific problems. For BGP traffic on IGW entities, such an outage can cause a large number of withdrawals to be exchanged between peers as routers experience path interruptions and some networks become unreachable. After a period of time, new routes will be advertised by the routers.

In case of anomalous network events, we start with NID as Step 1 following the proposed approach, where the BGP \hl{NI} datasets are used in \hl{\mbox{\cite{Al-Musawi2017}}}. Then, we conduct the network fault detection and localization as Step 2 through additional BGP \hl{NA} datasets available on the IGW. In Step 1, the datasets have 37 features with an output with four labels, i.e., Other (0) and Code Red I (1), Nimda (2), and Slammer(3), where the labels 1-3 indicate some well-known cyberattack incidents, and the label 0 represents the possible normal traffic or additional anomalous types of outputs to be processed in Step 2. In Step 2, we employ the BGP datasets on the edge routers in order to further explore the outputs, where there are two link failures considered in our datasets: one is between R1 and R2 on N1 and the other is between R5 and R6 on N2. Through the system analysis, the root cause analysis of the link outages can be narrowed down to the network interfaces on R1/R2 and R5/R6, respectively, using the system-specific datasets for identifying the root causes. The results of such a system reliability analysis step can lead to responsive and automated repair efforts of hardware or software issues. Based on our approach, we should note that a repair process does not necessarily affect network access. We can use the IGW entity on R2 or R9 to commission a HAP to provide a temporary link between the BS and satellite. 

\begin{table}
\centering
\renewcommand{\arraystretch}{1.2}
\caption{Accuracy and F1-Score of Mainstream ML Models}
\begin{tabular}{lcccc}
    \toprule
    Model & \multicolumn{2}{c}{Step 1} & \multicolumn{2}{c}{Step 2} \\
    & Accuracy & F1-Score 
    & Accuracy & F1-Score \\
     \midrule
    NB & 0.749 & 0.770 & 0.950 & 0.783\\
    \hline
    BN & 0.801 & 0.775 & 0.904 & 0.847\\
    \hline
    LR & 0.795 & 0.755 & 0.951 & 0.940\\
    \hline
    DT & 0.771 & 0.774 & \textbf{0.967} & \textbf{0.961}\\
    \hline
    RF & \textbf{0.839} & \textbf{0.821} & \textbf{0.970} & \textbf{0.961} \\
    \hline
    KNN &  0.807 & 0.797 & 0.962 & 0.952 \\
    \hline
    SVM & 0.781 & 0.692 & 0.932 & 0.905 \\
    \hline
    QDA & 0.756 & 0.696 & 0.943 & 0.928\\
    \hline
    LSTM & \textbf{0.835} & \textbf{0.813} & 0.959 & 0.956 \\
    \hline
    GRU & \textbf{0.834} & \textbf{0.811} & \textbf{0.963} & \textbf{0.963}\\
    \hline
    BLS & 0.825 & 0.799 & 0.959 & 0.937\\
    \hline
    XGBoost & \textbf{0.853} & \textbf{0.843} & \textbf{0.966} & \textbf{0.964}\\
    \bottomrule
\end{tabular}
\label{ml_results}
\end{table}

\begin{figure}[!ht]
\centering
\includegraphics[width=0.88\linewidth]{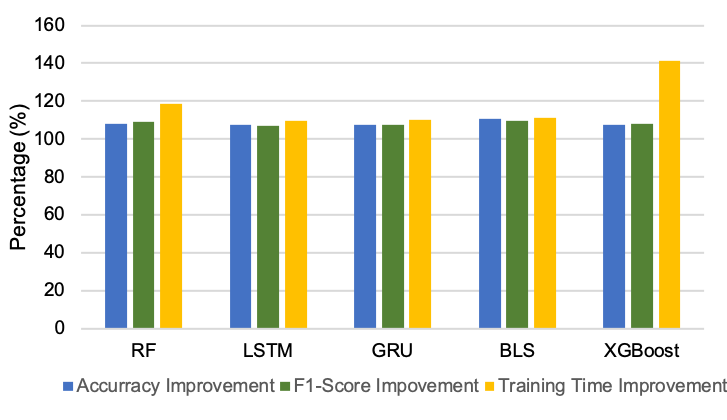}
\caption{ Performance improvements in ensemble and RNN methods}
\label{Fig:ml_comp}
\end{figure}

Now let us apply the mainstream ML methods (as listed in Table \ref{ml_results}) including the recently proposed NN-based algorithms to solve a multi-class classification problem in Steps 1 and 2. Naive Bayes \hl{(NB)} is a basic probabilistic classifier \hl{which} assumes the independence of input variables. \hl{Bayesian Network (BN)} is an algorithm that can solve classification problems based on the posterior probability of each class given the features. The logistic regression \hl{(LR)} and quadratic discriminant analysis (QDA) are parametric algorithms that can solve a classification problem. The decision tree \hl{(DT)} is a classical nonparametric algorithm for solving classification problems, while it sometimes suffers from the over-fitting problem. \hl{RF} is a popular ensemble method that can resolve the over-fitting issue. Support vector machines (SVM) and k-Nearest Neighbors (KNN) are nonparametric classification algorithms that have been broadly used in the literature. Long-short term memory (LSTM) is a special kind of recurrent \hl{NN (RNN), a powerful NN} for classification problems structured with input, hidden and output layers of neurons. LSTM addresses the gradient exploding problem of RNN and the gated recurrent unit (GRU) solves the gradient vanishing problem of LSTM. Broad Learning System (BLS) is a recently proposed \hl{method based on an improved random vector functional link NN.} \hl{The LSTM and GRU models are designed in a similar architecture: the first layer is an LSTM/GRU layer, followed by a fully connected layer with a `tanh' activation function and neurons equal to the dense units, and the last layer with a `softmax' activation function. Between these layers, two dropout layers are applied to avoid overfitting of the model.}

\hl{
Each model in Table \mbox{\ref{ml_results}} was trained on the datasets for two steps where the training sets are set to 60\%, and the average accuracy and F1-Score values are obtained on test sets. We tend to compare the ML algorithms with the same datasets with the aforementioned preprocessing from BGP raw data with one exception for QDA, where the Fisher score was used to reduce the features to 12 on the NI dataset due to its assumption on the covariance matrix for each class. We have extensively performed hyperparameter tuning in the grid search method for most ML algorithms, and we also used the popular AutoML tools, TPOT and Keras Tuner for hyperparameter tuning in applicable ML algorithms. The XGBoost model was tuned based on the results from TPOT. Due to the limited support for NN algorithms in TPOT, the Keras Tuner was used to tune hyperparameters in LSTM and GRU, where the optimal values of hyperparameters such as units and learning rates were searched in 200 epochs. The number of neighbors in KNN is set to 6 and 3 in Steps 1 and 2, respectively, while the number of estimators for RF is set to 200 and 60. For XGBoost, the maximum depth and minimum child weight are set to \{3, 1\} and \{1, 3\} with 100 estimators. For LSTM and GRU models, the hidden nodes, dense units and learning rates in Step 1 are set to \{180, 80, 0.0001\} and \{150, 200, 0.0001\}, respectively, followed by \{40, 180, 0.001\} and \{190, 120, 0.001\} in Step 2. For BLS, the `maptimes' and `enhencetimes' are set to \{5, 5\} and \{20, 50\} in Steps 1 and 2, respectively.
}

Table {\ref{ml_results}} shows the XGBoost, GRU, and RF have the best overall performance, while LSTM still has consistently good performance, followed by BLS in comparison to other ML methods. DT only has good performance in Step 2 but has under 77.5\% accuracy and F1-Score in Step 1. Fig. {\ref{Fig:ml_comp}} shows the top-performing ML methods can achieve improvements in training time, F1-Score, and accuracy, compared with the non-hierarchical approach, where fault detection is based on the combined datasets in one shot. With the proposed approach, BLS has the most significant improvement in accuracy to 111\%, followed by GRU and RF (108.1\%), LSTM (107.6\%), and XGBoost (107.4\%). The F1-Score is improved in BLS to 109.5\%, followed by RF (109\%), XGBoost (107.9\%), GRU (107.3\%), and LSTM (106.7\%). XGBoost leads the improvement in training time efficiency to 141.1\%), followed by RF (118.6\%), BLS (111\%), GRU (110.1\%) and LSTM (109.4\%). The results indicate the RNN methods (GRU and LSTM) and ensemble methods (XGBoost and RF) perform anomaly identification effectively.

\section{Conclusion}

This article discusses an ML-based hierarchical approach to self-maintenance for SICNs. The proposed approach aims to fill in the important management gap and facilitate increasing efforts on broadband CNs and data-driven autonomous management. Although this approach makes it an attractive topic to CN deployments and design, it also leaves much room for further research contributions from different perspectives, such as data collection or generation models, ML method enhancements, and evaluation on deployment variations.

\section*{Acknowledgment}
This work was supported by the High-Throughput and Secure Networks Challenge Program through the National Research Council of Canada. We thank Dr. Yeying Zhu for useful discussions and thank Huiqing Huang and Maham Bhatti for assistance in experimentation and data generation.

\ifCLASSOPTIONcaptionsoff
  \newpage
\fi

\begin{IEEEbiographynophoto}
{Peng Hu}
received his Ph.D. degree in Electrical Engineering from Queen's University, Canada. He is currently a Research Officer at the National Research Council of Canada, and Adjunct Professor at the University of Waterloo. 
\end{IEEEbiographynophoto}






\end{document}